# Automated placement of stereotactic injections using a laser scan of the skull

Margaret Henderson, Vadim Pinskiy, Alexander Tolpygo,
Stephen Savoia, Pascal Grange, Partha Mitra

*Cold Spring Harbor Laboratory*

## Abstract

Stereotactic targeting is a commonly used technique for performing injections in the brains of mice and other animals. The most common method for targeting stereoscopic injections uses the skull indentations bregma and lambda as reference points and is limited in its precision by factors such as skull curvature and individual variation, as well as an incomplete correspondence between skull landmarks and brain locations. In this software tool, a 3D laser scan of the mouse skull is taken *in vitro* and registered onto a reference skull using a point cloud matching algorithm, and the parameters of the transformation are used to position a glass pipette to place tracer injections. The software was capable of registering sample skulls with less than 100 micron error, and was able to target an injection in a mouse with error of roughly 500 microns. These results indicate that using skull scan registration has the potential to be widely applicable in automating stereotactic targeting of tracer injections.

## Introduction

Many techniques in modern experimental neuroscience involve performing procedures such as stimulation, electroporation, or injection on fine structures in the brain during surgery on a live animal. The most common technique for performing these procedures is stereotactic surgery. During stereotactic surgery, the animal is fixed in a stereotactic apparatus, the surface of the skull is exposed, and features of the skull are used to determine where to inject to reach a target brain location. Fixing the skull within the stereotactic frame places the brain within an arbitrary three-dimensional coordinate system, and the assumption is then made that this coordinate system corresponds to a reference atlas of the brain which can be used to target injections.

One important application of stereotactic surgery is in performing injections of tracer and other compounds for brain mapping. The Mouse Brain Architecture project is one such application, which makes use of stereotactic injections of tracers to map the connections between known areas of the mouse brain, with the goal of generating a comprehensive map of the mouse brain on the mesoscale of resolution (Bohland et al., 2009). Each mouse receives a single injection at one of 250 sites throughout the left hemisphere of the brain; the distribution of injection sites is based on the Allen Brain Atlas or ABA (Lein et al., 2009), with each site intended to target a defined brain area. After a transport period appropriate to the tracer used, the animal is perfused, the brain is frozen, sectioned, and stained, with alternating sections processed for conventional cell body staining and tracer detection through histochemistry or fluorescent imaging. The two-dimensional sections are then reconstructed computationally to generate a three-dimensional volume which is registered to an existing atlas in order to determine the final localization of the tracer within the brain, which provides information about the projections to and from the targeted brain area.

In the typical injection protocol used in the literature and also employed in the MBA project, stereotactic injections are made by locating the skull landmark bregma, located at the intersection of the coronal and sagittal sutures, and defining this as the origin of the stereotactic coordinate frame. A second landmark, lambda, is used to define the rostrocaudal axis, and a pair of ear bars is used to level the mediolateral axis. Once the animal is fixed within the stereotaxic frame, an injection site can be targeted by a manipulator in three dimensions relative to the location of bregma.

The use of bregma and other skull landmarks as reference points introduces opportunities for operator error and limits the precision of stereotactic injections. Since bregma serves as the zero point based on which the entire stereotactic coordinate system is defined, any inaccuracy in locating bregma has the potential to decrease the accuracy of injections significantly. Variability in locating bregma on the skull can arise from bias or error in individual technicians, as well as from differences in techniques for identifying bregma. One study found that using an algorithm to automate the location of bregma on the skull was able to improve the accuracy of injections, which indicates that at least some of the bregma-related error in injection targeting could be removed by perfecting the technique for locating bregma on the skull (Blasiak et al., 2002). However, even if bregma could be identified with perfect consistency on every mouse skull, some variability would persist in the accuracy of bregma-targeted injections due to the lack of an exact, predictable correspondence between the bregma point and the brain itself (Wahlsten et al., 1975; Chan et al., 2007).

The long-term goal of this project is to develop a method for stereotactic targeting in mice that does not use bregma exclusively as a reference point, in order to remove some of the error associated with bregma and make it possible to perform injections with greater precision. This can be accomplished by taking a laser scan of an individual mouse skull *in vivo* and registering it onto a reference skull using a surface matching algorithm. Once the skull is registered onto the reference, the parameters of the transformation can be used to obtain a corrected injection site.

**Methods**

The software package was designed using Matlab software using the GUI development environment (GUIDE). It utilized a CT scan of a skull with 25 μm voxel size, which was obtained from the Henkelman lab at the Hospital for Sick Children in Toronto, Canada. All skull scan data used in testing the GUI were obtained from male mice of strain C57BL/6J, of age 53±4 days and weight 18.8-26.4 grams. At the start of the procedure, mice were anaesthetized and mounted in the stereotactic frame, and the skull was exposed by making an incision down the midline of the scalp. All procedures for anaesthesia and surgery were in accordance with the guidelines of IACUC for CSHL. Before the scan was taken, the technician manually identified the skull landmarks bregma and lambda, as well as the bounding box for the scan, which defined an area of roughly 5 mm by 5 mm. A preexisting MATLAB GUI was used to take a scan of the exposed skull surface using a Keyence CCD laser displacement sensor. Scans ranged in resolution from 25x25 (625 points) to 250x250 (62,500 points).

Once the scan was complete, the sample skull was filtered for outliers, and aligned roughly to the reference skull using the location of bregma as identified by the technician. The symmetry plane of the data set through bregma was then determined, and this was used to determine y-axis and z-axis rotation (theta and phi). The symmetry plane was also used to determine the sagittal crest line of the data set, which was aligned to the sagittal crest line of the reference based on eikonal distance as an error metric. A three-dimensional grid was obtained for the reference skull, with each voxel containing the eikonal distance to the reference skull at that point, and this was used to obtain the eikonal distance for each point in the fitted sample. Registration of the crest line defined the x-axis rotation (nu) as well as translations in the y and z directions. Finally, dilation in the x, y, and z directions was determined by minimizing average eikonal distance over the entire point cloud.

Following alignment of the skulls, the user could enter an injection site in coordinates determined by the ABA, and receive a corrected set of coordinates on the sample skull relative to bregma. To calculate the injection site, the inverse of the original transformation was performed. To determine the effectiveness of the registration component of the software, average eikonal distance was calculated before fitting (after filtering for major outliers) and after fitting with the software. This was performed for five data sets.

The overall functionality of the software was tested on a single mouse by performing 4 injections using the skull registration algorithm. The typical protocol for injections uses a pressure gauge to manually level the skull before injecting in order to align the skull with the XY plane, and the automated protocol aims to improve on this step by detecting any rotations in the skull scan. To test this feature, two injections were made on a skull that had been leveled using the manual method and two were made after applying a known rotation. All injections were made with CTB and were visualized using fluorescence microscopy.

**Results**

As shown in Figure 1, the automated software ran successfully on multiple data sets and performed registrations. Average eikonal distance was used to measure the accuracy of the fit (Table 1). The average eikonal distance was lower for all data sets after fitting. Before registration (after filtering for outliers) the average eikonal distance was 265.76 microns, with a wide range of values from 124 microns to 420 microns. After fitting with the automated software, the values for average eikonal error ranged from 67 microns to 129 microns, with an average of 89.09 microns. The decrease in average eikonal distance before and after fitting with the software was statistically significant ($p < 0.05$, paired t-test).

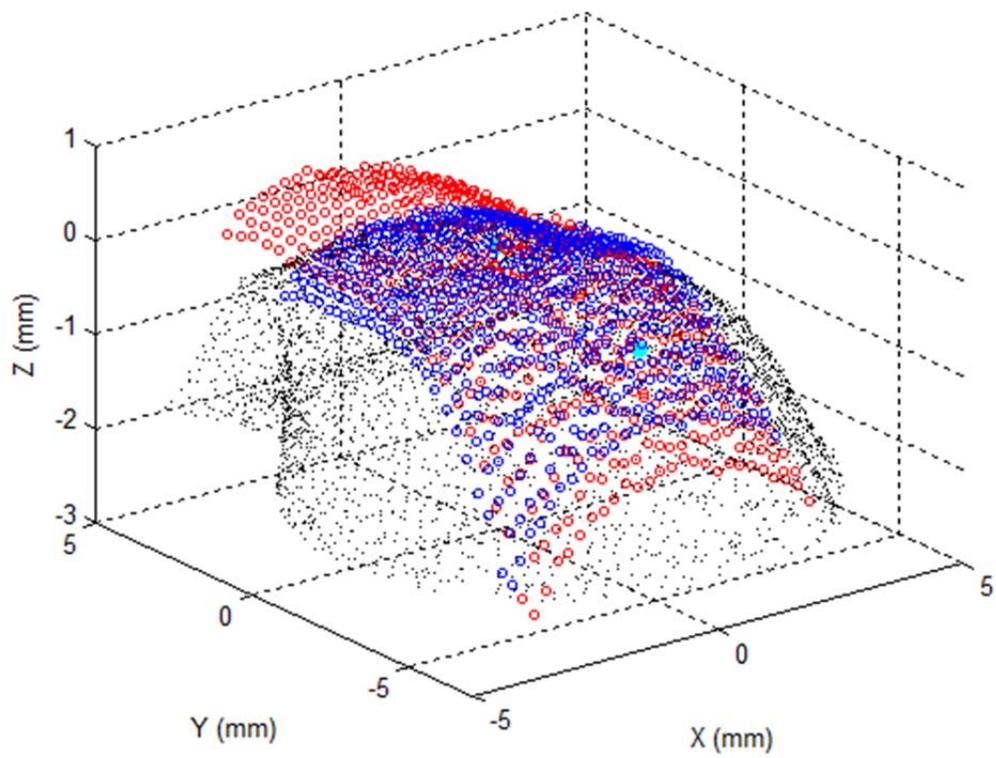

a.

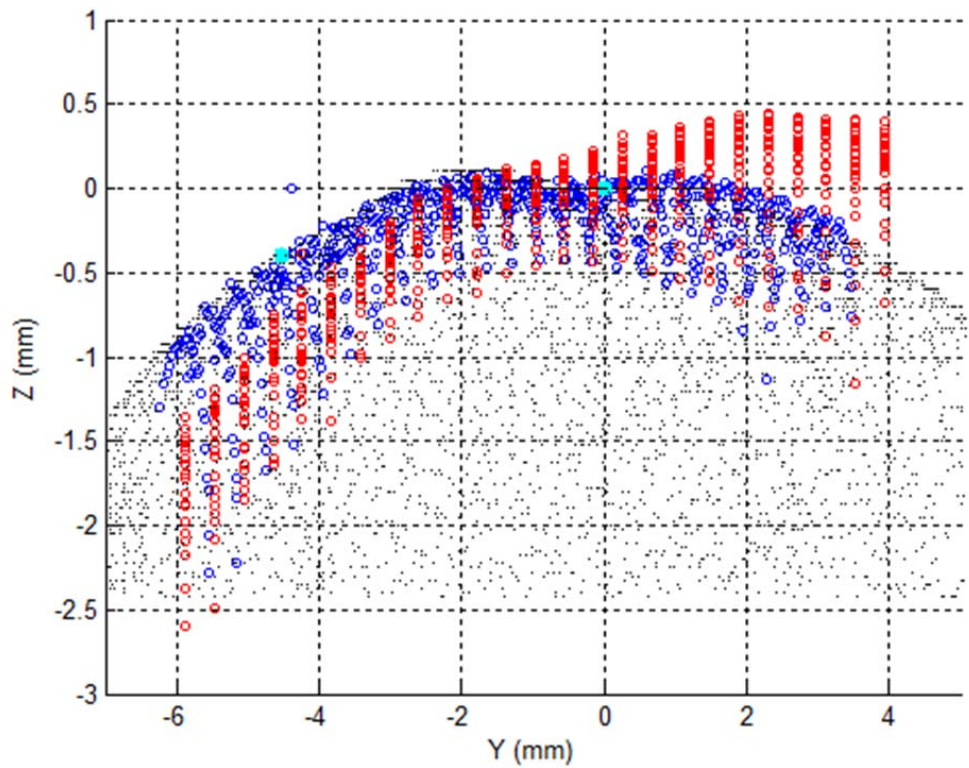

b.

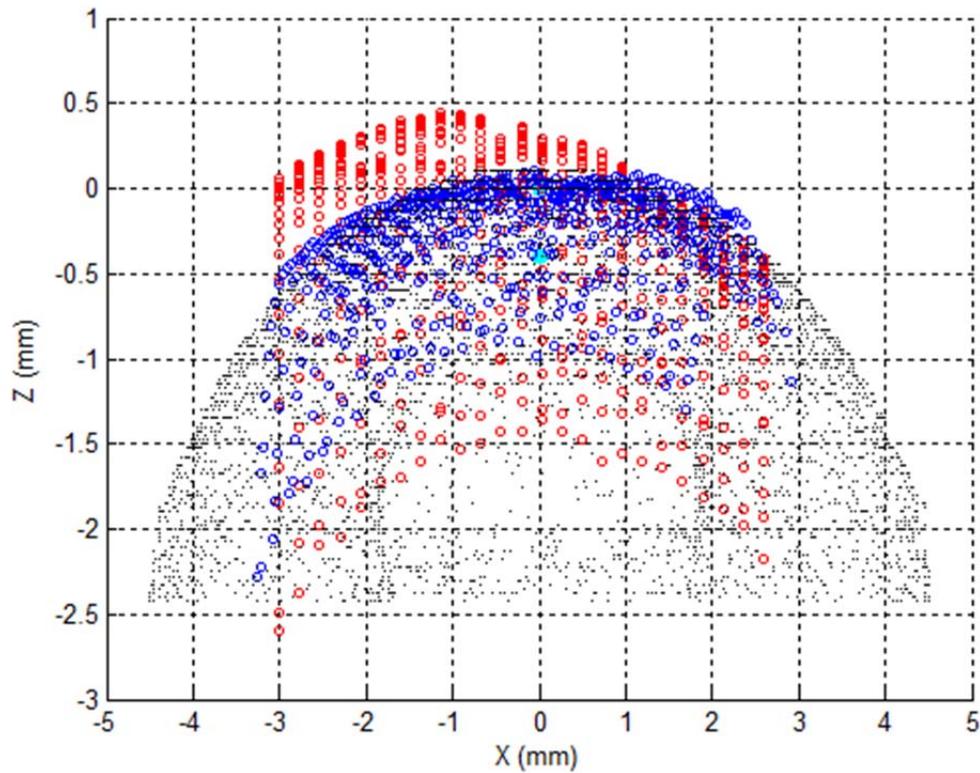

c.

**Figure 1.** A sample data set of 625 points from PTM681 was fitted to the reference skull using the automated software interface. CT scan is in black, original points in red, registered points in blue, bregma and lambda in cyan. **(a)** 3D view. **(b)** Sagittal view. **(c)** Coronal view.

**Table 1.** Average eikonal distance was calculated for 5 skull scan data sets (5,625 points each) before and after fitting to the reference skull using the registration software. For two data sets, fitting was also performed using the GUI interface as a comparison.

| Data Set | Before registration | After software registration |
|----------|---------------------|------------------------------|
| PTM649 | 420.29 | 129.03 |
| PTM656 | 168.43 | 43.48 |
| PTM658 | 235.25 | 102.72 |
| PTM680 | 124.16 | 91.65 |
| PTM681 | 380.66 | 78.57 |
| **μ** | **265.76** | **89.09** |
| σ | 129.94 | 31.54 |
| p (paired t-test) | | 0.0275 |

The software was tested and was successfully able to give a corrected injection position based on a set of input coordinates. Two injections were visualized using fluorescence microscopy to determine the actual sites where the injections landed (Table 2). The actual injection site was offset from the ABA coordinates by 0.53 mm in the left hemisphere and by 0.28 mm in the right hemisphere.

**Table 2.** Target sites were determined using the Allen Brain Atlas, and actual injection sites were determined using fluorescence microscopy after injections had been made using the automated software. All values are in mm.

Left hemisphere

|  | Target site | Actual injection site |
|---|---|---|
| x | -2.20 | -2.10 |
| y | -3.00 | -3.46 |
| z | 3.50 | 3.75 |
| Distance: | | **0.53** |

Right hemisphere

|  | Target Site | Actual injection site |
|---|---|---|
| x | 2.20 | 2.20 |
| y | -3.00 | -3.28 |
| z | 3.50 | 3.50 |
| Distance: | | **0.28** |

**Discussion**

The above results of the software test demonstrated that the software was functional and was able to target an injection with low to moderate error. The accuracy of the skull scan registrations as measured by the eikonal distance was on the order of 100 microns. These results have several implications for this project in the long-term. One important result is the confirmation that two point clouds representing the skulls of different individual mice can be registered together with fairly high accuracy using linear transformations. This indicates that if it is possible to determine the parameters for accurate registration between a given sample and a reference, implementing these transformations should be able to correct for individual skull variability and improve injection site accuracy. An important future step in this project will be to determine the level of registration accuracy needed for use in planning injections, and subsequently to improve the automated registration algorithm to reach this level of accuracy.

Another important factor in the effectiveness of this protocol will be the degree of correspondence between the reference skull used for registration and the reference brain atlas used to map injections. Using skull curvature to target injections relies on the assumption that the curvature and shape of an individual mouse skull is a good predictor of the shape, size and location of the underlying brain structures. Several studies have demonstrated that the mouse brain and skull develop in a concurrent and coordinated manner, leading to a tight fit of the brain within the skull and a high degree of correspondence between skull geometry and brain geometry in the adult mouse (Neiman et al., 2012; Boughner et al., 2008; Lierberman et al., 2008). This indicates that the mouse skull is closely aligned to the mouse brain *in-vivo*. However, the ABA is based on histological and cell body staining of a mouse brain post-perfusion, after extraction from the skull. Extracted mouse brains are subject to tissue deformation and shrinkage, and are also likely to be bent from their original orientation when removed from the physical support of the skull (Chan et al., 2007), meaning that the annotated brain found in the ABA cannot be assumed to correspond perfectly to an *in-vivo* mouse brain.

An important next stage of this project, therefore, will be to develop a brain atlas which more accurately represents the shape and size of the *in-vivo* mouse brain. This can be accomplished by performing a registration between the ABA and a scan of a mouse brain taken using MRI. The CT skull scan used in this project has associated with it an MRI scan from the same individual directly after excision of the brain from the skull. This high-resolution MRI was previously registered to a low-resolution MRI taken prior to extraction, so it can be assumed to be an accurate representation of the *in-vivo* brain shape and position relative to the skull. Therefore, performing a complete registration between the ABA and this MRI scan would subsequently result in an accurate registration between the CT reference skull and the ABA, making it possible to effectively use registration onto the reference skull to target injections accurately.

Improving the precision of stereotactic injections in mice could have significant implications to the MBA project as well as other experiments which use stereotactic injections. Currently, when an injection misses its target brain area, subsequent processing steps result in the determination of the actual injection site, and the correct area can be eventually reached by further injections. Reducing the number of off-target injections could potentially decrease the cost and increase the efficiency of projects that require many injections to be made. These applications are not limited to mice; stereotactic surgery is often used on other rodents such as rats and naked mole rats, and similar variability is encountered with the skulls of these animals (Xiao, 2006; Kline & Reid, 1984). If an appropriate atlas were available, it would be fairly simple to adapt this protocol to be used for stereotactic targeting in similar animals.

Developing a protocol for stereotactic targeting using skull curvature has significant potential to increase the precision of stereotactic injections, and could have numerous applications in experimental neuroscience. This project has confirmed that two mouse skulls can be registered together with reasonable accuracy using a series of linear transformations, and that a corrected injection site can be calculated using the parameters of this transformation. Progress has also been made in developing a method to automate the process of scanning and registration, which is an important step in making it possible to implement this protocol during stereotactic surgeries.

## Acknowledgements

We gratefully acknowledge support from the Mathers Foundation as well as NIH grant MH087988. We would like to acknowledge the CSHL undergraduate research program and the Joan Redmond Read Fellowship.